\documentclass[showpacs,preprintnumbers,preprint,amsmath,amssymb,nofootinbib]{revtex4}
\usepackage{dcolumn}    
\usepackage{latexsym}
\usepackage[dvips]{graphicx}
\usepackage{bm}

\newtheorem{prop}{Proposition}
\begin{document}
%

\title{\bf New instability in relativistic cylindrically symmetric system
}

\author{Ken-ichi Nakao}
\email{knakao@sci.osaka-cu.ac.jp}  
\author{Daisuke Ida}
\email{daisuke.ida@gakushuin.ac.jp} 
\author{Yasunari Kurita} 
\email{kurita@sci.osaka-cu.ac.jp}

\affiliation{ 
  $^{\ast}$Department of Mathematics and Physics, Graduate School of
  Science, 
   Osaka City  University, Osaka 558-8585, Japan.\\ 
  $^\dagger$ Department of Physics, 
   Gakushuin University, Tokyo 171-8588, Japan. \\
  $^\ddagger$ Osaka City University Advanced Mathematical Institute, 
  	Osaka 558-8585, Japan.
}

\begin{abstract}
We investigate an infinitesimally thin cylindrical shell composed of
counter-rotating dust particles. This system was studied by
Apostolatos and Thorne 
in terms of the C-energy for a bounded domain. 
In this paper, we reanalyze this system by evaluating the 
C-energy on the future null infinity.
We find that some class of 
momentarily static and radiation-free initial data does not settle
down into static, equilibrium configurations, and otherwise infinite
amount of the gravitational radiation is emitted to the future null infinity.
Our result implies the existence of an instability in this system.  
In the framework of the Newtonian gravity, a cylindrical shell
composed of counter-rotating dust particles
can be in a steady state with oscillation by
the gravitational attraction and centrifugal repulsion, and hence a static state
is not necessarily realized as a final state. By contrast, 
in the framework of general relativity, the 
steady oscillating state will be impossible since the gravitational
radiation will carry the energy of the oscillation to infinity. 
Thus, this instability has no counterpart in the Newtonian gravity. 

\end{abstract}


\preprint{ OCU-PHYS-281}
\preprint{ AP-GR-49}
\maketitle

\section{Introduction}
\label{sec:intro}

In order to understand the relativistic gravitational phenomena,
the existence of isometries is often or usually assumed, 
since the Einstein equations are very complicated 
system of quasi-linear partial differential equations. 
The simplest but useful assumption in relativistic astrophysical 
situations is that of spherical symmetry.
A shortcoming of this
assumption is that there is no freedom 
of gravitational radiation. 
It is hardly possible to extract
any effects of gravitational radiation within spherically symmetric systems.
In contrast, although the cylindrical symmetry might rarely appear in 
relativistic astrophysical situations,  
it has a degree of freedom of gravitational 
radiation known as Einstein-Rosen gravitational 
waves \cite{ER-wave}. This system has been studied 
in connection with the gravitational waves \cite{Marder,Stachel,Piran}. 

The cylindrically symmetric system has been investigated also in
connection with the hoop conjecture which states that 
{\it black holes with horizons form when and 
only when a mass M gets compacted into a region whose circumference $C$ in every 
direction is $C\lesssim 4\pi M$} \cite{Thorne:1972}. 
This conjecture might come from the well known 
difference between the spherically symmetric and the
cylindrically symmetric systems; the spherical gravitational collapse 
forms a horizon, whereas the cylindrical gravitational collapse 
does not, if energy conditions on 
material fields
are satisfied \cite{Thorne:1972,Hayward}. If the hoop conjecture 
is correct, the following statement also holds:
{\it if mass M does not get compacted in some direction, there is no horizon}. 
This means that if the spacetime singularity
forms but its mass $M$ does not get compacted in some direction, the 
spacetime singularity will be naked. 
Thus this conjecture is deeply related also 
to the cosmic censorship \cite{Penrose:1969}. 
In order to confirm the hoop conjecture, 
Shapiro and Teukolsky performed a numerical simulation for the
gravitational collapse of dust matter with spindle-like 
distributions \cite{Shapiro:1991a},
which is a relativistic counterpart
of Lin-Mestel-Shu collapse \cite{LMS}. 
Their result implies that a spacetime singularity without
a horizon forms from highly elongated distribution of matter.
This is a strong candidate for a counter-example 
of the cosmic censorship conjecture. 
However, based on the study about an infinitesimally thin cylindrical 
shell composed of dust particles with non-vanishing angular momenta, 
Apostolatos and Thorne argued that the effect of the rotation will halt the spindle 
gravitational collapse \cite{AT}.
However, any complete answer for this issue has not been given yet\cite{Shapiro:1992,SST}.  

In this paper, we reanalyze the cylindrical shell model originally studied by 
Apostolatos and Thorne. Our study is based on the 
evaluation of the C-energy at the future null infinity, whereas
the argument by Apostolatos and Thorne is on the C-energy for bounded domains.
Due to the difference between ours and theirs, the conclusion is  
rather different from each other. 

In this paper, 
we adopt the geometrized units $c=1=G$, and 
our notation follows the textbook 
of Hawking and Ellis \cite{Hawking-Ellis;1973}.

\section{Cylindrical shell composed of counter-rotating particles} 
\label{section:null}

\subsection{Spacetime with whole cylinder symmetry}

In this section, we review the infinitesimally thin cylindrical shell model studied by 
Apostolatos and Thorne\cite{AT}. Hereafter, we 
refer to this shell as the  
Apostolatos-Thorne (AT)-shell. 
The AT-shell is composed of dust particles with an identical rest mass
and an absolute value of angular momentum, and 
the half of those has a positive angular momentum, whereas the other half has 
negative one, such that the net angular momentum is zero,
and all the particles remain to be on a cylindrical shell. 
In this case, 
the spacetime $(M,\bm{g})$ has the whole cylinder symmetry or equivalently has the
metric tensor with the local 
form \cite{Melvin-1,Melvin-2}
\begin{equation}
\bm{g} = e^{2(\gamma^*-\psi)}(-dt^* \otimes dt^*+dr^*\otimes dr^*)
+r^2e^{-2\psi}d\varphi\otimes d\varphi
+e^{2\psi}dz\otimes dz,  
\end{equation} 
where $\gamma^*$, $\psi$ and $r$ depend on $t^*$ and $r^*$.
The ranges of coordinates are given by 
$-\infty<t^*,z<+\infty$, $0\le r^*<+\infty$, $0\le \varphi <2\pi$.
The axis of rotational symmetry is located at $r^*=0$.
The coordinate basis $\partial/\partial\varphi$ and 
$\partial/\partial z$ are the rotational and translational 
Killing vectors, respectively. 

If the stress-energy tensor, ${\bf T}$, satisfies the condition
\begin{eqnarray}
-{\bf T}\left({\partial\over \partial t^*},{\partial\over \partial t^*}\right)+
{\bf T}\left({\partial\over \partial r^*},{\partial\over \partial r^*}\right)=0,
\end{eqnarray}
then the Einstein equations impose 
the wave equation in the 2-dimensional Minkowski spacetime
on the radial coordinate function as
\begin{equation}
\partial_{t^*}^2r-\partial_{r^*}^2r=0.
\end{equation}
The solution of the above equation takes a form
\begin{equation}
r=f(v^*)+g(w^*),
\end{equation}
where 
\begin{eqnarray}
v^*=t^*+r^*,\quad
w^*=t^*-r^*,
\end{eqnarray}
and $f$ and $g$ are arbitrary functions. Here we
restrict ourselves to the case that 
$r=[{\rm const}]$ hypersurface is timelike, i.e., the inequality
$(\partial_{t*}r)^2<(\partial_{r*}r)^2$ holds. 
Then we adopt the metric variable, $r$, as a new radial coordinate,
and further adopt a function defined by 
\begin{equation}
t=f(v^*)-g(w^*)
\end{equation}
as the new time coordinate. 
The metric in this 
new coordinate system can be expressed as
\begin{equation}
\bm{g} = e^{2(\gamma-\psi)}(-dt\otimes dt+dr\otimes dr)
+r^2e^{-2\psi}d\varphi\otimes d\varphi
+e^{2\psi}dz\otimes dz, 
\label{metric} 
\end{equation} 
where
\begin{equation}
\gamma=\gamma^*-\frac{1}{2}\ln\left[(\partial_{r^*}r)^2-(\partial_{t^*}r)^2\right]
\end{equation}
has been defined.
Here only a pair of metric variables, $\gamma$ and $\psi$, appears. 

We assume that the AT-shell is put in the vacuum spacetime. 
Therefore, the metric 
takes the form of Eq.~(\ref{metric}) 
both inside and outside regions of the AT-shell. 
The Einstein equations lead to the equations for $\gamma$ and $\psi$ as
\begin{eqnarray}
&&\partial_t\gamma=2r(\partial_t\psi)\partial_r\psi, \label{Einstein-1}\\
&&\partial_r\gamma=r\left[(\partial_t\psi)^2+(\partial_r\psi)^2\right], 
\label{Einstein-2}\\
&&\left(\partial_t^2-\partial_r^2-\frac{1}{r}\partial_r\right)\psi=0.
\label{Einstein-3}
\end{eqnarray}
We also assume that the space is not closed in $r$-direction, i.e., 
\begin{equation}
re^\psi>0~~~~~~{\rm for}~~r>0.
\end{equation}

\subsection{Description of the AT-shell}

The trajectory of an AT-shell in the spacetime 
is a timelike hypersurface, $\Sigma_{\rm AT}$.
Though the spacetime is singular on $\Sigma_{\rm AT}$, it can consistently be treated
by Israel's metric junction method\cite{Israel-1,Israel-2,Israel-3}.  
The AT-shell, $\Sigma_{\rm AT}$,
divides the spacetime into two regions. 
We refer to the inside region of $\Sigma_{\rm AT}$ as $M_{-}$ 
and the outside one as $M_{+}$. 
Even if $\Sigma_{\rm AT}$ is singular, we can require 
that the metric tensor 
$\bm{g}$ and the Killing 
vectors $\partial/\partial\varphi$, $\partial/\partial z$ are continuous at 
$\Sigma_{\rm AT}$. It can be easily seen that 
the continuity of $\bm{g}(\partial/\partial\varphi, \partial/\partial\varphi)$ and 
$\bm{g}(\partial/\partial z, \partial/\partial z)$ 
implies the continuity of the coordinate function, $r$, and the 
metric variable, $\psi$, 
across $\Sigma_{\rm AT}$. By contrast, the continuity of $\gamma$
across $\Sigma_{\rm AT}$ is not guaranteed and this means that 
the coordinate function, $t$, and accordingly, the coordinate basis, $\partial/\partial t$,
may not be continuous across $\Sigma_{\rm AT}$. 
Then, the evolution of the cylindrical AT-shell is characterized 
by its radial coordinate $r=R(\tau)$, where $\tau$ is the proper time
naturally defined on the AT-shell. The circumferential radius of the AT-shell is given by  
${\cal R}(\tau)=e^{-\psi_{\rm s}(\tau)}R(\tau)$, where $\psi_{\rm s}(\tau)$ 
is the value of $\psi$ evaluated on the AT-shell.

We introduce the proper reference frame of an observer riding 
on the AT-shell as follows, 
\begin{eqnarray}
{\bf E}_U&=&X_{\pm}\frac{\partial}{\partial t_{\pm}}+V\frac{\partial}{\partial r}
={\rm four~velocity~of~the~shell}, \\
{\bf E}_N&=&V\frac{\partial}{\partial t_{\pm}}+X_{\pm}\frac{\partial}{\partial r}
={\rm outward~unit~vector~normal~to~the~shell}, \\
{\bf E}_\varphi&=&\frac{e^{\psi_{\rm s}}}{r }\frac{\partial}{\partial \varphi}, \\
{\bf E}_z&=&e^{-\psi_{\rm s}}\frac{\partial}{\partial z},
\end{eqnarray}
where
\begin{eqnarray}
V&:=&\frac{dR}{d\tau}, \\
X_{\pm}&:=&\frac{dt_{\pm}}{d\tau}=\sqrt{e^{-2(\gamma_{\pm}-\psi_{\rm s})}+V^2}.
\end{eqnarray}
The subscripts $+$ and $-$ are used to denote quantities evaluated on the outer and 
inner faces of the AT-shell, respectively, if necessary. 

As mentioned above,
the AT-shell is made of counter rotating dust particles 
which move along timelike geodesics whose tangents are denoted by ${\bf u}$. 
By virtue of the rotational isometry generated by $\partial/\partial\varphi$, 
the specific angular momentum (the angular momentum per unit rest mass) 
of each particle is conserved. 
Therefore, the component of ${\bf u}$ 
in the direction tangent to ${\bf E}_\varphi$
is given by
\begin{equation}
\bm{g}({\bf u},{\bf E}_\varphi)=\pm\frac{\alpha}{\cal R}=:\pm u,
\end{equation}
where $\alpha>0$ is a positive constant corresponding to an 
absolute value of the specific angular momentum. 
Since the rest mass of each particle is conserved quantity, the shell's rest 
mass per unit Killing length is also conserved, 
which we denote by $\lambda$.
We assume that
$\lambda$ is positive, {\it i.e.}, $\lambda > 0$ holds. Then the 
surface stress-energy tensor, ${\bf S}$, of the AT-shell is given by
\begin{equation}
{\bf S}=\sigma \left({\bf E}_U\otimes {\bf E}_U
+\frac{u^2}{1+u^2}{\bf E}_\phi\otimes{\bf E}_\phi\right),
\end{equation}
where  we have defined
\begin{equation}
\sigma:=\frac{\lambda\sqrt{1+u^2}}{2\pi R}
\end{equation}
as the surface energy density of the AT-shell.

In accordance with the Israel's prescription, the Einstein equations
for the AT-shell reduce to 
\begin{equation}
{\bf K}_{+}-{\bf K}_{-}
=8\pi \left[{\bf S}-\frac{1}{2}({\rm Tr}~{\bf S})\bm{h}\right],
\end{equation}
where ${\bf K}_+$ and ${\bf K}_-$ are the extrinsic curvatures of 
the AT-shell relative to the external region $M_+$ and 
the internal region $M_-$, respectively.
The above equation leads to
the junction conditions on the metric variables as
\begin{eqnarray}
&&{\bf E}_N\psi_{+}-{\bf E}_N\psi_{-}=-\frac{2\lambda}{R\sqrt{1+u^2}}, 
\label{junction-1}\\
&&X_{+}-X_{-}=-4\lambda\sqrt{1+u^2}, \label{junction-2}
\end{eqnarray}
and 
\begin{eqnarray}
\frac{d V}{d\tau}=V{\bf E}_U\psi_{\rm s}
-R\left[({\bf E}_U\psi_{\rm s})^2+({\bf E}_N\psi_{-})^2\right]
+\frac{X_{-}{\bf E}_N\psi_{-}}{1+u^2}
-\frac{X_{-}\lambda}{R(1+u^2)^{3/2}}+\frac{X_{-}X_{+}u^2}{R(1+u^2)},
\label{junction-3}
\end{eqnarray}
where
\begin{equation}
{\bf E}_Uf_{\rm s}=X_{\pm}\frac{\partial f_{\rm s}}{\partial t_{\pm}}
+V\frac{\partial f_{\rm s}}{\partial r}
~~~~~{\rm and}~~~~~
{\bf E}_Nf_{\pm}=V\frac{\partial f_{\pm}}{\partial t_{\pm}}
+X_{\pm}\frac{\partial f_{\pm}}{\partial r}.
\end{equation}

\subsection{Momentarily Static and Radiation-Free Initial Data}

Here we consider the initial data with the momentarily static 
and radiation-free (MSRF) conditions 
\begin{equation}
V=0~~~{\rm and}~~~\partial_t\psi=0=\partial_t^2\psi.
\end{equation} 
By solving the Einstein equations (\ref{Einstein-1})--(\ref{Einstein-3}), 
the metric variables in $M_{+}$ are given by 
\begin{eqnarray}
\gamma&=&\gamma_{+}+\kappa^2\ln(r/R), \\
\psi&=&\psi_{\rm i}-\kappa\ln(r/R),
\end{eqnarray}
where $\gamma_{+}$, $\psi_{\rm i}$ and $\kappa$ 
are integration constants. 
In contrast, the solutions in $M_{-}$ 
should satisfy the regularity condition at the symmetry axis, $r=0$, as
\begin{equation}
\gamma|_{r=0}=0,\quad\partial_r\psi|_{r=0}=0.
\end{equation} 
Since $\psi$ should be continuous across the AT-shell, 
the metric variables in $M_{-}$ are given by
\begin{equation}
\gamma=0,\quad\psi=\psi_{\rm i}. 
\end{equation}
The junction conditions (\ref{junction-1}) and (\ref{junction-2})
give relationships between 
the integration constants $\gamma_{+}$, $\kappa$, $\psi_{\rm i}$ and
the quantities characterizing the AT-shell as
\begin{eqnarray}
\gamma_{+}&=&-\ln(1-4\Lambda\sqrt{1+u^2}), \label{gamma+}\\
\kappa&=&\frac{2\Lambda}{(1-4\Lambda\sqrt{1+u^2})\sqrt{1+u^2}}, \label{kappa}
\end{eqnarray}
where $\Lambda:=\lambda e^{-\psi_{\rm i}}$ is the rest mass per unit
proper length. Here note that in order that the space is not closed in
$r$-direction, 
$\Lambda\sqrt{1+u^2}<1/4$ must be satisfied. 
This condition guarantees the positivity of $\kappa$, {\it i.e.}, $\kappa>0$. 

As shown by Apostolatos and Thorne, Eq.~(\ref{junction-3}) is rewritten
in the form
\begin{equation}
\frac{dV}{d\tau}=({\rm positive~quantity})\times \left[\Lambda_{\rm
    eq}-\Lambda\right],
\end{equation}
where
\begin{equation}
\Lambda_{\rm eq}(u):=\frac{u^2\sqrt{1+u^2}}{(1+2u^2)^2}.
\end{equation}
The above equation shows that if the rest mass per proper length
$\Lambda$ is greater than $\Lambda_{\rm eq}$, the MSRF AT-shell starts 
contracting, whereas if $\Lambda$ is smaller than $\Lambda_{\rm eq}$, 
it starts expanding. The initial data of $\Lambda=\Lambda_{\rm eq}$ 
corresponds to the static configuration. We can easily see from Eq.~(\ref{kappa}) 
that $\kappa=2u^2$ in the static case.

\section{C-energy argument for
 the final configuration}
\label{section:Final-config}

\subsection{C-energy}

The C-energy $E$ has been proposed by Thorne as a quasi-local energy 
which is the energy included within the cylinder with finite radius
and with unit Killing length \cite{Thorne:1965}.  
\begin{equation}
E:=\frac{1}{4}\left[\gamma^*
-\frac{1}{2}\ln\left\{(\partial_{r^*}r)^2-(\partial_{t^*}r)^2\right\}\right]
\end{equation}
Because the external region of the AT-shell is vacuum, 
the expression for the C-energy reduces to 
\begin{equation}
E:=\frac{\gamma}{4}. \label{E-def}
\end{equation}
As shown in the preceding section, the metric function, $\gamma$, diverges logarithmically
in the limit of
$r\to +\infty$ for the MSRF initial data. 
This means that  
$E(t,+\infty)$ of MSRF initial data is necessarily infinite
and thus the value of the ``total'' energy per unit Killing length is 
meaningless. The similar situations also appear  in the Newtonian gravity; 
the depth of gravitational potential produced by cylindrically 
distributed matter is infinite. 
However, the temporal variation of the total energy per unit Killing
length is meaningful and crucial for a later discussion. 

The advanced and retarded time coordinates $v$ and $w$, defined by
\begin{equation}
v=t+r,\quad w=t-r,
\end{equation}
are convenient to express
the C-energy carried by the gravitational waves. 
We refer to the null hypersurface, $v\to +\infty$, at infinity
as the future null infinity, ${\cal I}^{+}$.  
(About general arguments on the future null infinity in the spacetime 
with a space-translation Killing field, 
see Refs.\cite{Ashtekar-1,Ashtekar-2} .) 
Due to the gravitational emissions, the C-energy will vary with $w$ 
on the null hypersurface, 
$v=[{\rm const}]$. 
The retarded time function $w$  
plays a role of a time function on the null hypersurface 
given by $v=[{\rm const}]$.
From Eqs.~(\ref{Einstein-1}) and (\ref{Einstein-2}), 
the derivative of $E$ with respect to $w$ becomes 
\begin{equation}
\partial_w E=\frac{1}{4}\partial_w\gamma
=-\frac{r}{2}(\partial_w\psi)^2\leq0. \label{del-E}
\end{equation}
Taking the limit of $v\to +\infty$, it can be seen that
the C-energy is non-increasing function on ${\cal I}^{+}$
with respect to $w$,
which was shown by Stachel by the argument in terms 
of the news function~\cite{Stachel}. 

Here we study the time variation of the 
C-energy on ${\cal I}^{+}$, which will be caused by 
the gravitational waves from the AT-shell initially in the MSRF state.
Let the MSRF initial Cauchy surface 
be located at $t=t_{\rm i}$ in terms both of the  external and
internal time coordinates,
and let us refer to this Cauchy surface by $\Sigma_{\rm i}$. 
The initial radial coordinate of the AT-shell, 
$\Sigma_{\rm AT}\cap \Sigma_{\rm i}$, is denoted by $R=R_{\rm i}$.
Then the causal future of the initial location of the AT shell, 
$J^{+}(\Sigma_{\rm AT}\cap\Sigma_{\rm i})$, is given by the condition, 
$v_{+}\geq v_{\rm i}$ and $w_{-}\geq w_{\rm i}$, where $v_{\rm i}$ and $w_{\rm i}$
are given by
\begin{eqnarray}
v_{\rm i}:=t_{\rm i}+R_{\rm i},\quad
w_{\rm i}:=t_{\rm i}-R_{\rm i}. 
\end{eqnarray}
The portion of the spacetime,  
$D^+(\Sigma_{\rm i})-J^{+}(\Sigma_{\rm AT}\cap\Sigma_{\rm i})$, 
remains static, while the region $J^{+}(\Sigma_{\rm AT}\cap\Sigma_{\rm i})$
will be dynamical due to the
gravitational waves generated by the motion of the AT-shell. 

Let us describe the evolution of $\psi$ on $M_+$ as
\begin{equation}
\psi=\psi_{\rm i}-\kappa\ln\frac{r}{R_{\rm i}}+\delta_\psi.
\end{equation}
By virtue of the linearity of Eq.~(\ref{Einstein-3}), $\delta_\psi$ 
is subject to 
the same equation as that for $\psi$ as
\begin{equation}
\left(\partial_t^2-\partial_r^2-\frac{1}{r}\partial_r\right)\delta_\psi=0.
\end{equation}
Because the region $D^+(\Sigma_{\rm i})- J^{+}(\Sigma_{\rm AT}\cap\Sigma_{\rm i})$ 
remains static, 
the solution in $M_{+}$ of our interest takes the form~\cite{Landau-Lifshitz}
\begin{equation}
\delta_\psi=\int_{-\infty}^{t-r}\frac{p(\xi)}{\sqrt{(t-\xi)^2-r^2}}d\xi
=\int_{-\infty}^w\frac{p(\xi)}{\sqrt{(v-\xi)(w-\xi)}}d\xi,
\end{equation}
where $p(\xi)$ is a function which 
vanishes for $\xi< w_{\rm i}$ and will decay for $\xi\rightarrow +\infty$.  
The derivative of $\delta_{\psi}$ 
with respect to $w$ is given by
\begin{eqnarray}
\partial_w\delta_\psi 
&=& 
\lim_{\epsilon\rightarrow 0}
\frac{1}{\epsilon}\left[
\int_{-\infty}^{w+\epsilon}\frac{p(\xi)}
{\sqrt{(v-\xi)(w+\epsilon-\xi)}}d\xi
-\int_{-\infty}^w\frac{p(\xi)}
{\sqrt{(v-\xi)(w-\xi)}}d\xi
\right] \nonumber \\
&=& 
\lim_{\epsilon\rightarrow 0}
\frac{1}{\epsilon}\left[
\int_{-\infty}^w\frac{p(\xi+\epsilon)}
{\sqrt{(v-\xi-\epsilon)(w-\xi)}}d\xi
-\int_{-\infty}^w\frac{p(\xi)}
{\sqrt{(v-\xi)(w-\xi)}}d\xi
\right] \nonumber \\
&=&
\int_{-\infty}^w\frac{1}{\sqrt{w-\xi}}
\frac{d}{d\xi}\left(\frac{p(\xi)}{\sqrt{v-\xi}}\right)d\xi.
\end{eqnarray}
Substituting the above results into Eq.~(\ref{del-E}), we have
\begin{equation}
\lim_{v\rightarrow+\infty}\partial_wE=-\frac{P^2(w)}{2}, \label{Power}
\end{equation}
where
\begin{equation}
P(w)=\int_{-\infty}^w\frac{1}{\sqrt{w-\xi}}
\frac{dp}{d\xi}(\xi)d\xi
\end{equation}
has been defined. 
Provided that $P(w)$ is finite,
the rate of change in $E$ 
is also finite on ${\cal I}^{+}$. 

In order to define the quasi-local energy finite even in the limit of $r\rightarrow\infty$ 
for the system with whole cylinder symmetry, Thorne introduced 
an alternative definition of the C-energy as
\begin{equation}
E^{\rm (new)}:=\frac{1}{8}(1-e^{-8E}).
\end{equation}
For the MSRF initial data, 
this new version of the C-energy always becomes 
$1/8$ in the limit of $r\rightarrow\infty$. For $t>t_{\rm i}$, 
$\gamma$ in $M_{+}$ is written in the form 
\begin{equation}
\gamma=\gamma_{+}+\kappa^2\ln\frac{r}{R_{\rm i}}+\delta_\gamma. \label{gamma-after}
\end{equation}
From Eqs.(\ref{E-def}) and (\ref{Power}), we find that $\delta_\gamma$ 
at ${\cal I}^{+}$ is determined by
\begin{equation}
\lim_{v\rightarrow+\infty}\partial_w\delta_\gamma=-2P^2(w),
\end{equation}
and therefore, we have
\begin{equation}
\lim_{v\rightarrow+\infty}\delta_\gamma
=-2\theta(w-w_{\rm i})\int_{w_{\rm i}}^wP^2(\xi)d\xi, \label{delta-gamma}
\end{equation}
where $\theta(x)$ is the Heaviside's step function. 
From the above results,
it can be seen that $\delta_\gamma$ is finite.
Therefore, we have
\begin{equation}
\lim_{v\rightarrow+\infty}\partial_w E^{\rm (new)} 
=\lim_{v\rightarrow+\infty}\left(\frac{2R_{\rm i}}{v-w}\right)^{2\kappa^2}
e^{-2(\gamma_{+}+\delta_\gamma)}\partial_wE
= 0.
\end{equation}
Thus, $E^{\rm (new)}$ is constant on ${\cal I}^{+}$ and 
this is useless for discussing how much energy is released 
into the future null infinity, ${\cal I}^{+}$. 
For this reason, we adopt the original definition for the C-energy, $E$. 

\subsection{Final static configuration}

The MSRF initial data with $\Lambda=\Lambda_{\rm i}\neq\Lambda_{\rm eq}$ 
dynamically evolves. At first glance, the system seems to
settle down into a static configuration 
with $\Lambda=\Lambda_{\rm eq}$ by emitting or absorbing the
gravitational waves. Thus at first, we assume that the final configuration is
static and consider the relation between MSRF initial data and the
final configuration.  

The C-energy of the initial configuration $E_{\rm i}$ 
and that of the final configuration $E_{\rm f}$ are given by 
\begin{eqnarray}
E_{\rm i}&=&\frac{1}{4}\left[-\ln\left(1-4\Lambda_{\rm i}\sqrt{1+u_{\rm i}^2}\right)
+{\kappa_{\rm i}^2}\ln\frac{r}{R_{\rm i}}\right], 
\label{C-initial}\\
E_{\rm f}&=&\frac{1}{4}\left[-\ln\left(1-4\Lambda_{\rm eq}(u_{\rm f})
\sqrt{1+u_{\rm f}^2}\right)
+\kappa_{\rm f}^2\ln\frac{r}{R_{\rm f}}\right]. \label{C-final}
\end{eqnarray}
Here and hereafter, quantities with a subscript 
``i''
denote these of initial data, whereas those with 
subscript ``f'' denote those of the final configuration. 
Note that the C-energy at $w=w_{\rm i}$ is given by $E=E_{\rm i}$. 
After the system settles down in a static configuration, 
the C-energy in the causal future 
of the static AT-shell is given by Eq.~(\ref{C-final}). 
Thus the difference between the initial C-energy at $w=w_{\rm i}$ and the
final one is estimated on ${\cal I}^{+}$ as 
\begin{eqnarray}
\Delta E
&=&\lim_{v\rightarrow+\infty}\int_{w_{\rm i}}^{+\infty}
\frac{\partial E}{\partial w}dw \nonumber \\
&=&\lim_{r\rightarrow+\infty}\frac{1}{4}\left[\ln\frac{1-4\Lambda_{\rm i}\sqrt{1+u_{\rm i}^2}}
{1-4\Lambda_{\rm eq}(u_{\rm f})\sqrt{1+u_{\rm f}^2}}
+(\kappa_{\rm f}^2-\kappa_{\rm i}^2)\ln{r}
-\kappa_{\rm f}^2\ln R_{\rm f}+\kappa_{\rm i}^2\ln R_{\rm i}\right].
\end{eqnarray}
We can see from the above equation that if $\kappa_{\rm f}^2$ is different from 
$\kappa_{\rm i}^2$, the energy difference $\Delta E$ is infinite.  
However, we can see from Eqs.~(\ref{gamma-after}) and (\ref{delta-gamma}) that  
$\delta_\gamma$ is finite and thus the coefficients, $\kappa$, in the
logarithmic term is unchanged
by the gravitational emissions, {\it i.e.}, $\kappa_i=\kappa_f=\kappa$ hold
(this was pointed out by Marder \cite{Marder}). 
This fact is very important. 
On this ground,
we can uniquely specify 
the static configuration realized from a given MSRF
initial data. 

By virtue of the freedom 
of the constant scaling,  we can assume that the 
initial value of $\psi_{\rm s}$, {\rm i.e.}, $\psi_{\rm i}$, is zero, 
without loss of generality. 
Then, the initial rest mass per unit proper length $\Lambda_{\rm i}$ equals $\lambda$. 
The final 
static configuration is characterized only by $u=u_{\rm f}$
and  $\psi_{\rm s}=\psi_{\rm f}$,
noting that the specific angular momentum $\alpha$ and the rest mass per unit Killing length 
$\lambda$ are conserved quantities. 
Since the final configuration is static, 
\begin{equation}
\kappa=2u_{\rm f}^2
\end{equation}
holds.
Substituting this relation into
Eq.~(\ref{kappa}), we obtain
\begin{equation}
u_{\rm f}=\sqrt{\frac{\lambda}{\left(1-4\lambda\sqrt{1+u_{\rm i}^2}\right)
\sqrt{1+u_{\rm i}^2}}}.
\label{u-final}
\end{equation}
The final values of ${\cal R}$ and $\psi_{\rm s}$ are given by using
$u_{\rm f}$ in the form, 
\begin{eqnarray}
{\cal R}_{\rm f}&=&\frac{\alpha}{u_{\rm f}}, \label{R-final}\\
\psi_{\rm f}&=&\ln\frac{\lambda}{\Lambda_{\rm eq}(u_{\rm f})}. \label{psi-final}
\end{eqnarray}
Substituting Eq.~(\ref{u-final}) into the above equations, 
the final value of ${\cal R}_{\rm f}$ and $\psi_{\rm f}$ 
can be evaluated as functions of $u_{\rm i}$. 

Here we prove the following proposition.
\begin{prop}
Consider the MSRF initial data of the AT-shell composed of 
dust particles with non-zero specific angular 
momenta, i.e., $\alpha>0$ holds. 
If gravitational emissions from the AT-shell per unit Killing length are finite, 
and if 
the initial circumferential radius ${\cal R}_{\rm i}$ of the AT-shell
is greater than 
the critical value given by
\begin{equation}
{\cal R}_{\rm c}(u_{\rm i};\alpha,\lambda)
:=\sqrt{\frac{\alpha^2}{\lambda}\left(1-4\lambda\sqrt{1+u_{\rm i}^2}\right)
\sqrt{1+u_{\rm i}^2}}, \label{eq:Rc}
\end{equation}
where $u_{\rm i}=\alpha/{\cal R}_{\rm i}$,
i.e., ${\cal R}_{\rm i}>{\cal R}_{\rm c}$ holds,
then the AT-shell does not settle down into a
static configuration. 
\end{prop}
{\it Proof}: Suppose that 
the 
MNRS initial data of the AT-shell 
with the circumferential
radius ${\cal R}={\cal R}_{\rm i}$ settles down into 
a static configuration with the circumferential radius ${\cal R}_{\rm f}$. 
Provided that the gravitational emissions from the AT-shell are finite, 
Eqs.~(\ref{u-final}) and (\ref{R-final}) 
imply that the final circumferential radius ${\cal R}_{\rm f}$ is equal 
to ${\cal R}_{\rm c}$. 
Since $\kappa$ remains constant, 
the energy difference $\Delta E$ takes the form 
\begin{equation}
\Delta E=\frac{1}{4}(A+B),
\end{equation}
where
\begin{eqnarray}
A&=&\ln\frac{1-4\lambda\sqrt{1+u_{\rm i}^2}}
{1-4\Lambda_{\rm eq}(u_{\rm f})\sqrt{1+u_{\rm f}^2}},\\
B&=&-2u_{\rm f}^2\ln\frac{R_{\rm f}}{R_{\rm i}}.
\end{eqnarray}
Using Eqs.(\ref{u-final}), we rewrite $A$ in the form
\begin{equation}
A=\ln\left[
1-\frac{4u_{\rm i}^4\hat{\lambda}(1-\hat{\lambda})}
{1+4u_{\rm i}^2(1+u_{\rm i}^2)(1-\hat{\lambda})}
\right], 
\label{A}
\end{equation}
where $\hat{\lambda}:=\lambda/\Lambda_{\rm eq}(u_{\rm i})$. 

The condition that the space is not closed in $r$-direction 
leads the following inequality:
\begin{equation}1-4\lambda\sqrt{1+u_{\rm i}^2}>0, \end{equation}
which is equivalent to
\begin{equation}
1+4u_{\rm i}^2(1+u_{\rm i}^2)(1-\hat{\lambda})>0.
\end{equation}
From Eq.~(\ref{A}) and the above inequality,
$A$ is positive if $\hat{\lambda}>1$ holds.
This condition, $\hat{\lambda}>1$, is equivalent to
the condition ${\cal R}_{\rm f}<{\cal R}_{\rm i}$,
which immediately follows from the equation
\begin{equation}
\frac{{\cal R}_{\rm f}^2}{{\cal R}_{\rm i}^2}
=\frac{u_{\rm i}^2}{u_{\rm f}^2}=1+\frac{(1+2u_{\rm i}^2)^2}{\hat{\lambda}}(1-\hat{\lambda}),
\label{key-eq}
\end{equation}
where Eq.~(\ref{u-final}) has been used.
Hence $A$ is positive for ${\cal R}_{\rm f} < {\cal R}_{\rm i}$. 

Next, from the definition of ${\cal R}$ and Eq.~(\ref{psi-final}), we have 
\begin{equation}
\frac{R_{\rm f}}{R_{\rm i}}=
\frac{\lambda{\cal R}_{\rm f}}{\Lambda_{\rm eq}(u_{\rm f}){\cal R}_{\rm i}}
=\hat{\lambda}\frac{u_{\rm i}\Lambda_{\rm eq}(u_{\rm i})}
{u_{\rm f}\Lambda_{\rm eq}(u_{\rm f})}. 
\label{B-2}
\end{equation}
From Eq.~(\ref{key-eq}), we have
\begin{equation}
\hat{\lambda}=\frac{4(1+u_{\rm i}^2)+1/u_{\rm i}^2}{4(1+u_{\rm i}^2)+1/u_{\rm f}^2}.
\end{equation}
Substituting the above equation into Eq.~(\ref{B-2}), we have
\begin{equation}
\frac{R_{\rm f}}{R_{\rm i}}=
\frac{4(1+u_{\rm i}^2)+1/u_{\rm i}^2}{4(1+u_{\rm i}^2)+1/u_{\rm f}^2}
\cdot
\frac{(1+2u_{\rm f}^2)^2}{(1+2u_{\rm i}^2)^2}.
\cdot
\frac{u_{\rm i}^3\sqrt{1+u_{\rm i}^2}}
{u_{\rm f}^3\sqrt{1+u_{\rm f}^2}}.
\end{equation}
It is easily seen from the above equation that $R_f = R_i$ holds,
if $u_{\rm f}=u_{\rm i}$ or equivalently 
${\cal R}_{\rm f}={\cal R}_{\rm i}$ holds. 
The partial derivative of 
$R_{\rm f}/R_{\rm i}$ with respect to $u_{\rm f}$ with $u_{\rm i}$ fixed 
becomes
\begin{eqnarray}
\frac{\partial}{\partial u_{\rm f}}
\left(\frac{R_{\rm f}}{R_{\rm i}}\right)
&=&-\frac{4(1+u_{\rm i}^2)+1/u_{\rm i}^2}{[4(1+u_{\rm i}^2)+1/u_{\rm f}^2]^2}
\cdot
\frac{1+2u_{\rm f}^2}{u_{\rm f}^6(1+u_{\rm f}^2)^{3/2}} 
\cdot\frac{u_{\rm i}^3\sqrt{1+u_{\rm i}^2}}
{(1+2u_{\rm i}^2)^2} \nonumber \\
&\times&\left[4(1+u_{\rm i}^2)u_{\rm f}^4+4(2+3u_{\rm i}^2)u_{\rm f}^2+1\right].
\end{eqnarray}
Therefore, $R_{\rm f}/R_{\rm i}$ is a decreasing function of $u_{\rm f}$
for fixed $u_{\rm i}$, 
in the domain, 
$u_{\rm i}>0$. 
This implies that
$R_{\rm f}< R_{\rm i}$ holds, 
if $u_{\rm f}  > u_{\rm i}$ or equivalently ${\cal R}_{\rm f} < {\cal R}_{\rm i}$ holds. 
Hence $B$ is positive, if ${\cal R}_{\rm f} < {\cal R}_{\rm i}$ is satisfied.

Combining the above results, it is concluded that
$\Delta E>0$ holds, {\it i.e.}, the C-energy on ${\cal I}^{+}$ 
increases, 
if ${\cal R}_{\rm i} > {\cal R}_{\rm f}={\cal R}_{\rm c}$ holds. 
However this is impossible, 
since the C-energy is non-increasing function on
${\cal I}^{+}$.  \hfill $\Box$

Finally, we prove the following proposition. 
\begin{prop}
There exists a MSRF initial data set of an AT-shell composed of 
dust particles with non-zero specific angular momenta, 
$\alpha>0$, which does not settle down into the static configuration. 
\end{prop}
\noindent
{\it Proof}: We shall show that the MSRF initial data with the condition 
${\cal R}_{\rm i}>{\cal R}_{\rm c}$ exists. 
The condition ${\cal R}_{\rm i}>{\cal R}_{\rm c}$ is equivalent to 
\begin{equation}
\frac{\lambda}{u_{\rm i}^2\left(1-4\lambda\sqrt{1+u_{\rm i}^2}\right)
\sqrt{1+u_{\rm i}^2}}>1. 
\end{equation}
We rewrite the above inequality in the form
\begin{equation}
F(u_{\rm i}^2)>0,  \label{F-inequality}
\end{equation}
where
\begin{equation}
F(x)=\lambda^2(2x+1)^4-x^2(1+x).
\end{equation}
In order that the space is not closed in $r$-direction, 
the inequality
\begin{equation}u_{\rm i}^2<{1\over 16\lambda^2}-1\end{equation}
must be satisfied. 
(In order that this condition is satisfied, $\lambda$ must be smaller
than 1/4.) Here we investigate whether Eq.~(\ref{F-inequality}) 
is possible in the domain $0<u_{\rm i}<\sqrt{1/16\lambda^2-1}$. 
It is easy to see 
\begin{equation}
\lim_{u_{\rm i}\rightarrow 0}F(u_{\rm i}^2)=\lambda^2>0.
\end{equation}
Since $F(x)$ is continuous, 
the intersection between   
the domain $0<u_{\rm i}<\sqrt{1/16\lambda^2-1}$ and the open 
neighborhood of $u_{\rm i}=0$ is not empty, 
in which the inequality (\ref{F-inequality}) is satisfied. \hfill $\Box$

\section{Summary and discussion}
\label{section:summary}

Apostolatos and Thorne showed that the C-energy within a bounded
domain $r<r_0$ of MSRF configuration is minimized by the equilibrium 
configuration with both $\alpha$ and $\lambda$ fixed and 
further with $\psi(r_0)=0$. They also proved that the 
C-energy within the domain $r<r_0$ of a dynamical configuration 
is always greater than that of a MSRF
configuration that has the same $\alpha$, $\lambda$, $R$ and
$\psi(r_0)=0$, but different $V$, $\psi(r)$ and $\partial_t\psi(r)$. 
By noticing that the 
gravitational radiation carries the C-energy from a bounded domain 
to outside of it, they inferred from the 
above two facts that the energy of the oscillating AT-shell is 
released by the gravitational emissions and then AT-shell 
will settle down into a static, equilibrium 
state. Their speculation seems to be physically reasonable, but  
rigorous proof has not been given there. 

In this paper, by assuming that gravitational radiation 
carries finite C-energy from the AT-shell to the future null
infinity, we have shown that the MSRF initial data
of the AT-shell does not necessarily settle down into the static state. 
Our argument is based on
the non-increasing nature of the C-energy on the future null
infinity. If the initial circumferential radius of the AT-shell 
is greater than that of the expected final static state, then 
the AT-shell can not settle down into the static state.  
By contrast, it is not forbidden by this C-energy
argument that the AT-shell with an initial 
circumferential radius smaller than the expected finial static value 
settles down to the equilibrium configuration. 
We expect from this fact that the outward motion  
of AT-shell is caused by gravitational emissions. 
If this expectation is real, the AT-shell with ${\cal R}_{\rm i}>{\cal R}_{\rm c}$ 
will go to infinity by the secular outward motion. 
Thus the static, equilibrium configuration of the AT-shell will be 
unstable for the outward displacement. 
Since this secular motion may be ascribed to gravitational radiation, 
there is no Newtonian counterpart of this instability. 

At first glance, this behavior of the AT-shell is somewhat terrible,
but it might not be so if we consider a cylinder of infinitesimally thin
shell composed of dust particles but with finite length $L$ which is
initially much greater than its circumferential radius ${\cal R}$. The motion
of this finite cylindrical thin shell will be well approximated by the 
AT-shell model. Then, due to the instability mentioned above, ${\cal R}$ of
the central part of the cylinder might become larger and larger, but finally
its motion cannot be approximated by the AT-shell when ${\cal R}$
becomes  comparable to $L$. In other words, the initial cylindrical shape
might approach to the spherical shape by the  gravitational emission. 
If this expectation is real, 
our situation seems to be similar to the well known phenomenon that 
the initial eccentric orbit of a compact star
binary approaches to the circular orbit by the gravitational
emissions\cite{Peters}. 
However, it has not been rigorously shown that the above scenario is true.
Rather, at present, there remains a possibility that the AT-shell 
collapses to form a naked singularity by this instability. In
order to see whether the scenario of outward secular motion is true,  we
have got to investigate directly the solution, but this is a future work.

\begin{acknowledgments}
It is our pleasure to thank Hideki Ishihara for his valuable discussion. 
We are also grateful to colleagues in the astrophysics and gravity 
group of Osaka City University for helpful discussion and criticism.
YK is supported by the 21st Century COE ``Constitution of wide-angle 
mathematical basis focused on knots'' 
from the Ministry of Education, Culture,
Sports, Science and Technology (MEXT) of Japan.
\end{acknowledgments}

\end{document}